\documentclass[12pt]{article}
 \def\gae{\; ^{>}_{\sim} \;}

\title{\textbf{Comments on SUSY inflation models on the brane}}
{\author{\\[1cm]
{\sc \large Lu-Yun Lee$^{1,\dag}$, Kingman Cheung$^{1,2,3,\ddag}$, and Chia-Min Lin$^{1,\S}$}\\
{\sl\small $^1$Department of Physics, National Tsing Hua University, Hsinchu, Taiwan 300 }\\
{\sl \small $^2$Division of Quantum Phases \& Devices, School of Physics,}\\
{\sl\small Konkuk University, Seoul 143-701, Republic of Korea} \\
{\sl\small $^3$Physics Division, National Center for Theoretical Sciences,
Hsinchu 300, Taiwan}
}}

\usepackage[margin=2cm]{geometry}
\usepackage{graphicx,color}
\usepackage{subfigure}
\begin{document}
\maketitle
\begin{abstract}
In this paper we consider a class of inflation models on the brane where the dominant part of the inflaton scalar potential does not depend on the inflaton field value during inflation. In particular, we consider supernatural inflation, its hilltop version, A-term inflation, and supersymmetric (SUSY) D- and F-term hybrid inflation on the brane. We show that the parameter space can be broadened, the inflation scale generally can be lowered, and still possible to have the spectral index $n_s=0.96$.

\end{abstract}
\footnoterule{\small $^\dag$d9522809@oz.nthu.edu.tw,
$^\ddag$cheung@phys.nthu.edu.tw, $^\S$cmlin@phys.nthu.edu.tw}
\section{Introduction}
\label{sec1}

It is well-known that if we consider a braneworld scenario where our four-dimensional world is viewed as a 3-brane embedded in a higher-dimensional bulk, the Friedmann equation can be modified into \cite{Cline:1999ts, Csaki:1999jh, Binetruy:1999ut, Binetruy:1999hy, Freese:2002sq, Freese:2002gv, Maartens:1999hf}
\begin{equation}
H^2=\frac{1}{3 M_P}\rho \left[1+\frac{\rho}{2\Lambda}\right],
\label{eq5}
\end{equation}
where $\Lambda$ may provide a relation between the four-dimensional Planck scale $M_4$ and five-dimensional one $M_5$ through
\begin{equation}
M_4=\sqrt{\frac{3}{4\pi}}\left(\frac{M_5^2}{\sqrt{\Lambda}}\right)M_5,
\end{equation}
and $M_P=M_4/\sqrt{8\pi} \simeq 2.4 \times 10^{18}\mbox{ GeV}$ is
the reduced Planck scale. We will set $M_P=1$ in the following. The
nucleosynthesis limit imples that $\Lambda \gae (1\mbox{ MeV})^4
\sim (10^{-21})^4$. A more stringent constraint, $M_5 \gae 10^5
\mbox{ TeV}$, can be obtained by requiring the theory to reduce to
Newtonian gravity on scales larger than $1\mbox{ mm}$
\cite{Maartens:1999hf}, this corresponds to $\Lambda \gae (10^{-16})^4$.
In this paper, we will consider some consequences of modified
Friedmann equation to inflationary cosmology. Let us start by
presenting some standard results from \cite{Maartens:1999hf}. The
slow-roll parameters are given by
\begin{eqnarray}
\epsilon & \equiv & \frac{1}{2} \left(\frac{V^\prime}{V}\right)^2 \frac{1}{\left(1+\frac{V}{2\Lambda}\right)^2} \left(1+\frac{V}{\Lambda}\right) \label{eq0},\\
\eta & \equiv & \left(\frac{V^{\prime\prime}}{V}\right)\left(\frac{1}{1+\frac{V}{2\Lambda}}\right).
\label{eq1}
\end{eqnarray}
The number of e-folds is
\begin{equation}
N=\int^{\phi(N=60)}_{\phi(N=0)} \left(\frac{V}{V^{\prime}}\right) \left(1+\frac{V}{2\Lambda}\right)d\phi.
\label{eq2}
\end{equation}
The spectrum is
\begin{equation}
P_R=\frac{1}{12\pi^2}\frac{V^3}{V^{\prime 2}}\left(1+\frac{V}{2\Lambda}\right)^3.
\label{eq3}
\end{equation}
The spectral index is
\begin{equation}
n_s=1+2\eta-6\epsilon.
\end{equation}
One may think that when the effect of being ``on the brane'' is
significant, i.e., $V/2\Lambda  \gg 1$,  the slow-roll parameters
approaches to zero and the spectral index will be very close to one.
This suggests that inflation on the brane is not favored by the
latest Wilkinson Microwave Anisotropy Probe (WMAP) result
($n_s \simeq 0.96$) \cite{Komatsu:2008hk}. However, we
show in this paper that it is not necessarily the case. For some models the
spectral index is independent of $V/2\Lambda$, while the spectral index
will be \textit{lowered} by increasing $V/2\Lambda$ in some other models.
This is the main result of this work.

In general Eq. (\ref{eq2}) is not easy to solve by hand. However, if
we have an inflation model with a potential $V \simeq V_0$ during
inflation, where $V_0$ is independent of the inflaton field $\phi$,
then from Eqs. (\ref{eq1}), (\ref{eq2}), and (\ref{eq3}) we found that
the Cosmic Microwave
Background (CMB) normalization, i.e., $P_R \simeq (5 \times
10^{-5})^2$ \cite{Komatsu:2008hk}, at $N=60$ requires $U$ to be a constant,
 which is defined as
\begin{equation}
U \equiv V_0 \left(1+\frac{V_0}{2\Lambda}\right).
\end{equation}
$U$ will be a constant with the value $U=V_0$ when $\Lambda
\rightarrow \infty$. When $V_0/2\Lambda$ is large, $V_0$ is small but
$U$ is unchanged in order to get the same $N$ and $P_R$. This implies
$\eta$ is unchanged also. Note that this trick does not apply to
Eq. (\ref{eq0}). Instead, when we fix $U$, we found that $\epsilon$
actually becomes larger by a factor of $V/\Lambda$! Here we have to be
careful about inflation which may be destroyed by large $\epsilon$,
therefore a lower bound of $\Lambda$ is expected. One may wonder for
such a large $\epsilon$, maybe it is possible to get a large tensor to
scalar ratio. However, the answer is negative because the tensor to
scalar ratio is exactly suppressed by a factor of $(1+V_0/\Lambda)$
\cite{Maartens:1999hf} which cancels the increase of $\epsilon$.

\section{Supernatural inflation on the brane}
\label{sec2}

Supernatural inflation \cite{Randall:1995dj} is a supersymmetric
version of tree level hybrid inflation with a potential of the form
during inflation:
\begin{equation}
V=V_0+\frac{1}{2}m^2\phi^2.
\end{equation}
This kind of potential on the brane was considered in
\cite{Bento:2002kp}. However, we are focusing on supernatural
inflation and we will use a different method to deal with it.

In conventional ($V/2\Lambda \rightarrow 0$) case, we have
\begin{equation}
\phi=\phi_e e^{\frac{Nm^2}{V_0}},
\end{equation}
and
\begin{equation}
P_R=\frac{1}{12\pi^2}\frac{V_0^3}{m^4 \phi^2}.
\end{equation}
The gist of supernatural inflation is that if we put $V_0=M_S^4$
where $M_S \sim 10^{-7} \sim 10^{11}\mbox{ GeV}$ is the
gravity-mediated supersymmetry breaking scale and $m \sim O(1)\mbox{
TeV}\sim 10^{-15}$ the soft mass. We can satisfy CMB normalization with $\phi_e \sim 10^{-8}$ and $N=60$.

Now in order to know what will happen if we have supernatural
inflation on the brane, as pointed out in Sec. \ref{sec1}, we do not
need to solve Eqs. (\ref{eq0})-(\ref{eq3}). We just demand
\begin{equation}
U=V_0 \left(1+\frac{V_0}{\Lambda}\right)=(10^{-7})^4,
\label{eq4}
\end{equation}
and we can satisfy CMB normalization with every parameters (namely,
$\phi$, $\phi_e$, $N$ and $m$) the same. Therefore we can see the
effect of being on the brane is that $V_0$ can be lowered. How low
could $V_0$ be can be found from our requirement $V \simeq V_0$ (or
$V_0 > (1/2)m^2\phi^2 \sim 10^{-46}$). So the lower bound can be
achieved by $\Lambda^{1/4} \sim 10^{-16}$ which coincides the lower bounds of $\Lambda$ mentioned in Sec. \ref{sec1}. A lower value of $V_0$
means we can also use, for example, gauge-mediated SUSY breaking \cite{Giudice:1998bp} to
build a supernatural inflation.

Let us estimate the value of $\epsilon$, originally we have $\epsilon=(1/2)m^4 \phi^2/V_0^2 \sim 10^{-20}$. Now it will increase by a factor of $V_0/\Lambda \sim 10^{18}$ so we have $\epsilon \sim 0.01$. On the other hand, $\eta=0.01$ is unchanged. Therefore the spectral index $n_s=1+2\eta-6\epsilon=0.96$! We have shown that for a supernatural inflation on the brane with a concave upward potential, $n_s=0.96$ may be achieved with lowest $\Lambda^{1/4}$. This can be tested by searching for large extra dimension.

\section{Hilltop inflation on the brane}
\label{sec4}
The spectral index for supernatural inflation will be increased with $\Lambda$, and eventually get larger than one. In this case a method to lower
the spectral index is provided in \cite{Lin:2009yt}. The idea was to
convert the model into a hilltop inflation \cite{Boubekeur:2005zm,
Kohri:2007gq}. In this case, the potential becomes
\begin{eqnarray}
V&=&V_0+\frac{1}{2}m^2\phi^2-\lambda \phi^4 \\
 &\equiv& V_0 \left(1+\frac{1}{2}\eta_0 \phi^2\right)-\lambda \phi^4,
\end{eqnarray}
where $\lambda \sim A/M_P \sim 10^{-15}$ is from an A-term and
$n_s=0.96$ can be naturally achieved. Again when we consider
inflation on the brane, we just need to replace $V_0$ by $U$ and in
this case we redefine $\eta_0 \equiv m^2/U$, then every parameter
will remain unchanged. Since in this case we have $\phi_e=10^{-9}$
\cite{Lin:2009yt}, the lower bound of $V_0$ can be reduced by two
orders or so. In this case, no matter how $V_0/2\Lambda$ changed, we
still get $n_s=0.96$. This can be seen from Eq. (\ref{eq1}), because
$U$ is fixed, although large $V/2\Lambda$ in the right parenthesis
trying to reduce $\eta$, $V_0$ will also reduce correspondingly and
$\eta$ remain fixed. In order to avoid large $\epsilon$ to reduce
the spectral index. We have to make sure $\epsilon \ll |\eta|$, but
it is no more difficult to achieve for hilltop inflation because
$\epsilon$ is further suppressed by the quartic term in the
potential.

\section{(Hybrid) A-term inflation on the brane}
\label{sec3}

Another kind of model with this nice property is A-term inflation.
We consider the potential of the form \cite{Allahverdi:2006ng, Lin:2009ux}
\begin{equation}
V=V_0+\frac{1}{2}m^2 \phi^2 -A \frac{\lambda_p \phi^p}{p M^{p-3}_P}+\lambda^2_p. \frac{\phi^{2(p-1)}}{M_P^{2(p-3)}}
\end{equation}
When $V_0=0$, it becomes MSSM inflation \cite{Allahverdi:2006iq, Allahverdi:2008zz, GarciaBellido:2006fd, Allahverdi:2006we} or A-term inflation \cite{Lyth:2006ec, Bueno Sanchez:2006xk, Allahverdi:2006cx}.
During inflation the dominant part of the scalar potential is given by
\begin{equation}
V(\phi_0)\sim m^2 \phi^2_0 +V_0,
\end{equation}
which is independent of $\phi$ and
\begin{equation}
\phi_0= \left(\frac{m M_P^{p-3}}{\lambda_p \sqrt{2p-2}}\right)^{1/(p-2)}
\end{equation}
is the field value at the saddle point where $V^\prime=V^{\prime\prime}=0$. It was shown in \cite{Allahverdi:2006wt} that the spectral index $0.92 \leq n_s \leq 1$ can be obtained depending on deviation from the saddle point condition.
As long as we make sure $\epsilon \ll |\eta|$ in this model, the spectral index is unchanged. We just set
\begin{equation}
U \equiv V(\phi_0) \left(1+\frac{V(\phi_0)}{\Lambda}\right),
\end{equation}
which is a constant and can be determined for any $V(\phi_0)$.
The constraint of $\lambda$ may be stronger in this kind of model, because $\epsilon$ is not much smaller than $\eta$, but since the scale of A-term inflation can be very low. We do not have to worry about large $V/\Lambda$.

\section{F- and D-term inflation on the brane}
D-term inflation on the brane has been considered in
\cite{Panotopoulos:2005hc}, and F-term case in
\cite{McDonald:2002bd, BoutalebJ:2003ge}. In \cite{McDonald:2002bd},
it was shown that the vacuum energy induced mass\footnote{This is
used to be called "Hubble induced mass", but clearly this name is
not appropriate in this context.}  can be suppressed hence evade the
notorious $\eta$-problem. This can be seen from Eq. (\ref{eq5}). We
can think the Hubble parameter is obtained from $U$ via $H^2=U/3$,
however the induced mass will be $m_{induced}^2 \sim V_0$. Since
$V_0<U$ the slow-roll condition $m^2<H^2$ is preserved.
Interestingly, the suppressed reduced mass is also crucial for
hilltop version of F-term inflation \cite{Lin:2008ys, Pallis:2009pq}
where the spectral index can be lowered to $n_s=0.96$. In
\cite{BoutalebJ:2003ge}, the authors got an upper bound for
$\Lambda$. However we think here is no upper bound for $\Lambda$ and
when $\Lambda \rightarrow \infty$, conventional case is recovered.
Instead, we think there may exist a lower bound for $\Lambda$.

For D-term inflation, it was shown in \cite{Panotopoulos:2005hc}
that the scale can be lowered (as expected). We think maybe it is
possible to solve the well-known cosmic string problem, however, we
do not know how the cosmic string bound will change in this set-up.
In \cite{Panotopoulos:2005hc}, the author think the spectral index
should be always very close to $1$ ($n_s=0.99$). However, based on
our argument, we think the conventional case $n_s=0.98$ can be
achieved\footnote{But anyway, $0.98$ and $0.99$ are close.} or even
lower $n_s$ may be generated through large $\epsilon$. We also have
a hilltop version of D-term inflation \cite{Lin:2006xta,
Lin:2007va}, which can reduce the spectral index to $n_s=0.96$ and
it will be independent of $V_0/2\Lambda$.
\section{Conclusions}
\label{5}
In this paper, we considered a class of supersymmetric inflation models in the framework of a braneworld. The criteria is the dominant part of the scalar potential during inflation is independent of the field value. We found the slow-roll parameter $\epsilon$ can become large and the spectral index can be lowered. The spectral index will be a constant if $\epsilon \ll |\eta|$ is satisfied. Generally the scale of inflation $V_0^{1/4}$ can be lowered as long as $V_0$ still dominates the potential and $\epsilon$ do not become too large to destroy inflation or make the spectral index too low. In particular $n_s=0.96$ can be obtained in many inflation models in the braneworld scenario.

\section*{Acknowledgement}
This work was supported in part by the NSC under grant No. NSC
96-2628-M-007-002-MY3, by the NCTS, and by the Boost Program of
NTHU, and by WCU program through the NRF funded by the MEST
(R31-2008-000-10057-0). CML would like to thank David H. Lyth, Anupam Mazumdar, and John McDonald for their helpful comments.

\end{document}